\documentclass[12pt]{article}
\usepackage[table,xcdraw]{xcolor}
\usepackage[ruled, linesnumbered, vlined]{algorithm2e}
\usepackage{amsmath,amsfonts,amssymb,graphicx,geometry,authblk,color,soul,comment,hyperref,mathtools,subcaption}  %svg
\title{ Learning the Effective Adhesive Properties of Heterogeneous Substrates}
\author{Maximo Cravero Baraja\footnote{This work was conducted at the California Institute of Technology while MCB was a Visiting Student Researcher from the \'Ecole Polytechnique F\'ed\'erale de Lausanne, Switzerland} }
\author{Kaushik Bhattacharya\footnote{Corresponding Author: bhatta@caltech.edu}}
\affil{California Institute of Technology}
\date{}

\begin{document}
\maketitle

\begin{abstract}

Adhesion is a fundamental phenomenon that plays a role in many engineering and biological applications.   This paper concerns the use of machine learning to characterize the effective adhesive properties when a thin film is peeled from a heterogeneous substrate.  There has been recent interest in the use of machine learning in multiscale modeling where macroscale constitutive relations are learnt from data gathered from repeated solution of the microscale problem.   We extend this approach to peeling; this is challenging because peeling from heterogenous substrates is characterized by pinning where the peel front gets stuck at a heterogeneity followed by an abrupt depinning.  This results in a heterogeneity dependent critical force and a singular peel force vs. overall peel rate relationship.   We propose a neural architecture that is able to accurately predict both the critical peel force and the singular nature of the peel force vs. overall peel rate relationship from the heterogeneous adhesive pattern.  Similar issues arise in other free boundary and free discontinuity problems, and the methods we develop are applicable in those contexts.

\end{abstract} 

%%%%%%%%%%%%%%%%%%%%%%%%%%%%%%
%%%%%%%%%%%%%%%%%%%%%%%%%%%%%%
\section{Introduction}
\label{ch:introduction}

Adhesion is a fundamental phenomenon that plays a role in many engineering and biological applications. The ability to understand and characterize adhesive properties of different systems is important for the development of wearable medical devices \cite{iqbal_mahgoub_du_leavitt_asghar_2021}, durable coatings \cite{joseph_rajput_patil_nisal_2021}, and advanced manufacturing techniques \cite{LEPOIVRE2020948}, among others. 

In many engineering materials, heterogeneity at the microscale plays a crucial role in macroscopic  properties \cite{10.1115/1.1483342}. The ability to link these fine-grained properties to macroscopic behavior is invaluable in designing or characterizing materials and structures. In this work, we investigate this micro-macro relationship in the context of peeling a thin film from a heterogeneous adhesive substrate. In doing so, we consider a propagating front impeded by obstacles. This formulation also lends itself to applications in crystal growth, phase boundaries, dislocations, and crack propagation \cite{10.2307/2030137}. 

The first model for the peeling of a thin film goes back to  Rivlin \cite{rivlin_effective_1944}, who derived a relationship between the effective work per unit area required to peel the film, and the applied peel force and angle.  This model assumed an inextensible tape and a brittle adhesive.  This model was generalized by Kaelble \cite{kaelble_theory_1959}, Gent and Hamed \cite{gent_hamed_peel_mechanics}, and Kendall \cite{kendall_peeling_solid_films} to other situations.   More recently, interest has turned to the role of heterogeneities and its effect on peeling \cite{ghatak_2004,xia_adhesion_2013, majumder_2012}.   In particular, Xia {\it et al.} \cite{xia_2012,xia_adhesion_2013,xia_adhesion_2015} investigated the driving factors influencing the effective adhesive strength of a thin film in two contexts. First, they studied the effects of heterogeneities in the elastic stiffness of the film, demonstrating significant improvements in the effective peeling force without modifying the underlying adhesive \cite{xia_adhesion_2013}. In a subsequent investigation, they studied the same properties but using a uniform thin film and a heterogeneous adhesive interface \cite{xia_adhesion_2015}. 
In this work, we  focus on the latter problem. In particular, we investigate a data-driven approximation to the relationship between heterogeneous microstructures and the overall adhesive properties.

The use of machine learning is emerging as an important methodological advancement in mechanics. The specific applications vary widely, but one promising line of work is multiscale modeling, where one seeks to learn the macroscale constitutive relation using data gathered from repeated solution of the microscale problem.  To elaborate, in multiscale modeling a unit cell problem (a model or partial differential equation at the small scale) has to be solved at each time step and each discretization point of the large scale problem.  A brute-force implementation of this approach (often called FE$^2$ or concurrent multiscale) is prohibitively computationally expensive and one seeks to use data gathered from repeated solutions of the small scale problem to train a neural network surrogate of the solution operator of unit cell problem.

There are a number of works on learning the solution operator of partial differential equations \cite{zhu_2018,bhattacharya_2021,geist_2021,lu_learning_2021,khoo_2021,kovachki_2023}.  There are also a number of works on modeling complex constitutive behaviors.   Wang \textit{et al.} \cite{WANG2018337} apply a hybrid data-driven approach combined with classical constitutive laws to study hydromechanical properties of porous media with pores at various scales. Vlassis \textit{et al.} \cite{VLASSIS2020113299} use graph neural networks (GNNs) as surrogates for anisotropic hyperelastic models in order to predict the homogenized behavior of polycrystals. Meyer \textit{et al.}  \cite{MEYER2022111175} used GNNs to compute structure-property relationships for a large family of shell-lattice topologies.  Convolutional neural networks have been used to predict yield surfaces by encoding information at the microscale in tomographic images by Heidenrich \textit{et al.\ } \cite{HEIDENREICH2023103506}.
% inverse design
% Bessa cites
% \cite{BESSA2017633}, \cite{Bessa_Glowacki_Houlder_2019}, \cite{doi:10.1073/pnas.1911815116} 
%Another relevant field of application is that of inverse design, namely using models that can predict performance metrics of a material or structure to optimize the input parameters. Bessa \textit{et al.}\ used Bayesian ML to design new metamaterials \cite{Bessa_Glowacki_Houlder_2019}, as well as proposing a general framework for data-driven material modeling and design \cite{BESSA2017633}.
Finally, there are a number of works on learning time-dependent behavior using recurrent neural networks (RNNs).  Ghavamian and Simone \cite{ghavamian_accelerating_2019} suggested that an RNN with long short-term memory (LSTM) can be used to describe effective plastic behavior of a representative volume and then be used in FE$^2$ calculations.  Mozaffar {\it et al.} \cite{mozaffar2019deep} showed that an RNN with gated recurrent units (GRUs) is able to approximate the homogenized behavior of a composite medium consisting of isotropic elastic inclusions in an isotropic rate-independent plastic matrix.  Wu and Noels \cite{wu_recurrent_2022} seek to  predict macroscopic response and microscopic strain distribution using a (GRU) RNN combined with PCA of the strain-field.  Liu {\it et al.} \cite{liu_learning_2023} introduced the recurrent neural operator and used it for multiscale modeling of polycrystalline plasticity.
 
In this paper, we explore the use of machine learning to predict the effective behavior of free boundary and free discontinuity problems in heterogeneous media.  Such problems are characterized by pinning -- where the moving boundary gets stuck at a heterogeneity followed by an abrupt depinning and a rapid advance (e.g., \cite{barabasi_1995}).  Consequently, the effective behavior is characterized by a critical force which separates a pinned or sessile regime where the boundary does not move, from an unpinned or glissile regime where the interface moves with an average velocity that depends on the applied force \cite{Dirr_Yip_2006,dondl_bhattacharya_2016}.  Further, the transition is abrupt and characterized by a singular behavior near the critical force.    All of this creates challenges in the learning problem and we develop methods to overcome these challenges.  We may consider peeling of an adhesive film from a heterogeneous surface as a typical system, and the approach here is applicable to other free boundary problems including crystal growth, phase boundaries and dislocations.   Fracture is a free discontinuity problem and more difficult, and we comment on it later.

We introduce the peeling problem in detail in Section \ref{sec:peeling}.  We recall the governing equations and the effective behavior, and then introduce the numerical methods in this section.  Section \ref{ch:ml_microstructure_characterization} describes the learning problems we consider, the data generation and the training.  We present the results in Section \ref{ch:results_characterization} and conclude with Section \ref{sec:conclusion}.

\section{Peeling Dynamics} \label{sec:peeling}

\subsection{Peeling} \label{ssec:peeling_model}
\begin{figure}
    \centering
    \includegraphics[width=3in]{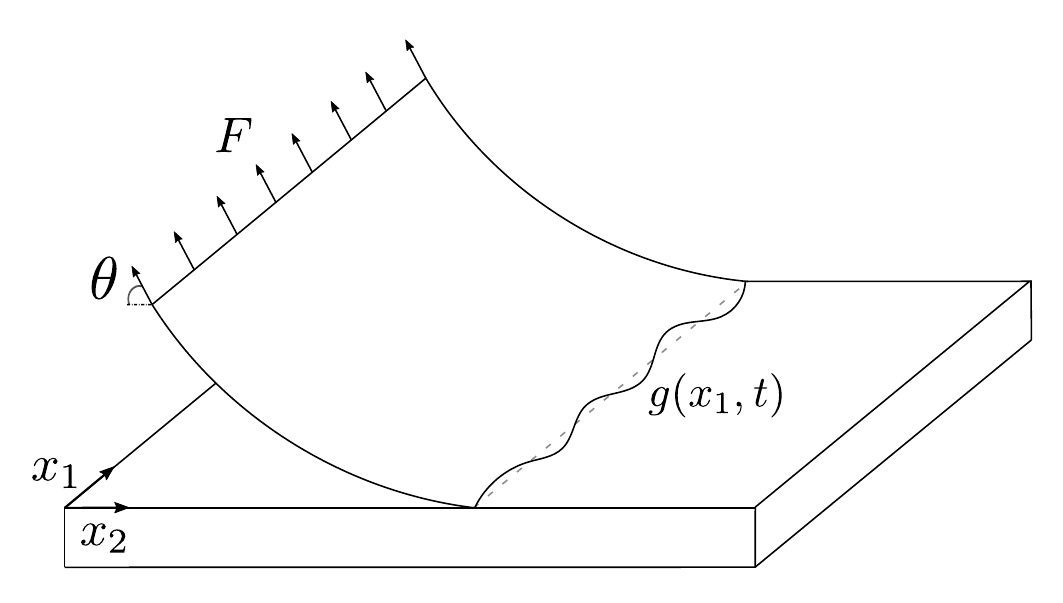}
    \caption{Thin film peeling configuration}
    \label{fig:peeling_setup}
\end{figure}

We consider the peeling of a thin film that is adhered to a rigid surface by the application of a force $F$ at an angle $\theta$ at the far edge as shown in Figure \ref{fig:peeling_setup}.  We assume that the adhesive strength is possibly heterogeneous, i.e. it depends on the position on the substrate.  Consequently the peel front is not necessarily straight: the peeling force tries to move the peel front forward, while the adhesive strength resists this motion.  Further, the non-straight peel front causes corrugations in the the unpeeled region of the thin film, and the energy associated with the bending favors a straight peel front.  Consequently the peel front at time $t$ may be described as a graph of a function $g$ with interface $\Gamma(t) = \{ x_2 = g(x_1,t)\}$ in the coordinate system shown in Figure \ref{fig:peeling_setup}. 

The peel front evolves according to the integro-differential equation \cite{xia_adhesion_2015}
\begin{align} \label{eq:peeling}
    g_{t}(x_{1},t) = \big(F(1 - \cos\theta)(1-4(-\Delta)^{1/2}g(x_{1},t)) - \varphi(x_{1}, g(x_{1},t)) \big)_+, \quad g(x_{1}, 0) = g_{0}(x_{1})
\end{align}
where $(q)_+ = \max\{q,0\}$ is the non-negative part of $q$, and the half Laplacian $(-\Delta)^{1/2}$ is given by
\begin{equation}
    (-\Delta)^{1/2}g(x_{1}, t) = \int_{-\infty}^{\infty}\frac{g(y, t) - g(x_{1}, t)}{|y-x_{1}|}dy.
\end{equation}
The equation states that the velocity of the peel front is the difference between the geometry-dependent driving force $F(1-\cos \theta)$ that drives the peel front forward\footnote{In what follows, the peeling angle is set to $\theta = \pi / 2$ and hence the factor $(1- \cos \theta)$ is dropped.  The results can be generalized to the case of arbitrary $\theta$ by redefining $F$ if necessary.}, and the resistance induced by the deviation from straight front due to the resistance to bending (the half Laplacian) as well as the adhesive strength.    Finally, a region of film that is peeled does not re-adhere or heal, and so the peel front can only advance.  This is why we restrict the peel-front velocity $g_t$ to be non-negative.  See Xia \textit{et al.} \cite{xia_adhesion_2015} for further details.  Note that is equation is both nonlinear (due to the dependence of $\varphi$ on $g$), and nonlocal (due to the half Laplacian).

We assume that the adhesive strength $\varphi$ is periodic.   After an initial transient depending on the initial condition $g_0$, the solution $g(x_1,t)$ becomes periodic in $x_1$ for each $t$ ($g(x_1,t) = g(x_1+\ell) \quad \forall \ t$ where $\ell$ is the period).   If the initial condition is periodic in $x_1$, it remains periodic.  So, we restrict ourselves to initial conditions $g_0$ and solutions $g$ that are periodic in $x_1$.

\begin{figure}
    \centering
    {\includegraphics[width=6in]{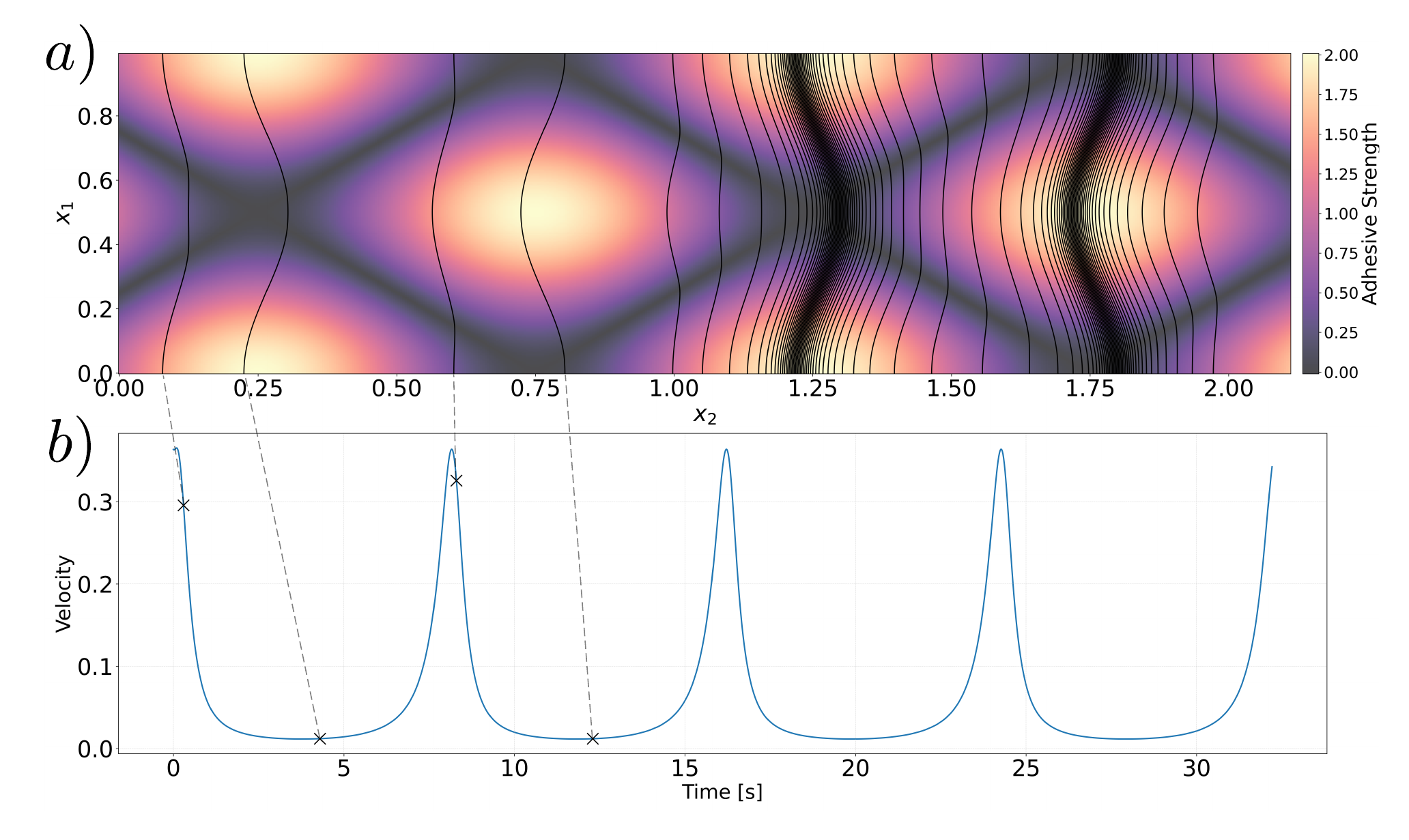}}
    \caption{A representative simulation of peeling subject to constant forcing $F=1$ and peeling angle $\theta = 90^{\circ}$.  (a): Adhesive strength distribution and snapshots of the peel front. (b): Spatially averaged velocity vs time.} % \hl{Mark a,b}}
    \label{fig:peel_front_evo}
\end{figure}

A representative result (obtained by a numerical method to be described later) is shown in \autoref{fig:peel_front_evo}. Figure \ref{fig:peel_front_evo}(a) depicts the spatial distribution of the adhesive strength $\varphi$ in color. The light and dark regions correspond to higher and lower adhesive strengths respectively, as indicated by the color bar. The black lines depict snapshots of the peel front.  These snapshots are taken at equally spaced time intervals with longer time intervals in the left half of the figure and shorter time intervals on the right half of the figure.  Figure \ref{fig:peel_front_evo}(b) depicts the spatially averaged velocities $\bar{v}(t) = \frac{1}{\ell}\int_{0}^{\ell}\dot{g}(x_{1}, t)dx_{1}$ at each point in time $t$ where $\ell$ is the periodicity.  The velocities corresponding to the snapshots of the peel front on the left region of the microstructure are indicated by a marker on the velocity curve. 

We see that the heterogeneities in the microstructure cause fluctuations in the velocity, which is also evident from the inconsistent spacing between the peel front snapshots.   Further, the snapshots associated with lower speeds occur either in the regions where the adhesive strength is high across the peel front or where the adhesive strength varies resulting in higher bending.  This is consistent with (\ref{eq:peeling}), where both bending and the adhesive strength lead to a reduction in the peel front velocity. On the other hand, we see a straighter peel front traveling at a higher speed in areas of uniformly lower adhesion.

\subsection{Effective behavior}

We are interested in the overall behavior for peeling over distances significantly larger than the periodicity. The representative result in Figure \ref{fig:peel_front_evo} shows that after an (short) initial transient, the peel front evolution becomes periodic.  Further, we observe that the peel front undergoes a {\it pulsating} motion (a superposition of periodic and steady motions) in $t$ where it advances through the unit period at a fixed interval of time
\begin{equation} \label{eq:pulsating}
g(x_1,t) = \frac{\ell}{T}t + g_\text{per}(x_1,t) \iff
g(x_1,t+T) = g(x_1,t) + \ell \quad \forall \ t \ \text{large enough}.
%g(x_1+\ell,t + T) = g(x_1, 
\end{equation}
We can use this pulsating motion to define an effective behavior.

Indeed, in a model without healing\footnote{We believe based on numerical evidence  that this result holds even with healing, but this has not been established rigorously.}, Dondl and Bhattacharya \cite{dondl_bhattacharya_2016} (following Yip and Dirr \cite{Dirr_Yip_2006} in a similar problem) show that the  the peeling front will always undergo a pulsating motion.  Specifically, given $F, \varphi$, there exists a unique $T>0$ independent of $x_1$ such that (\ref{eq:pulsating}) holds for some $g_\text{per}$. 

\begin{figure}
    \centering
    \includegraphics[width=4in]{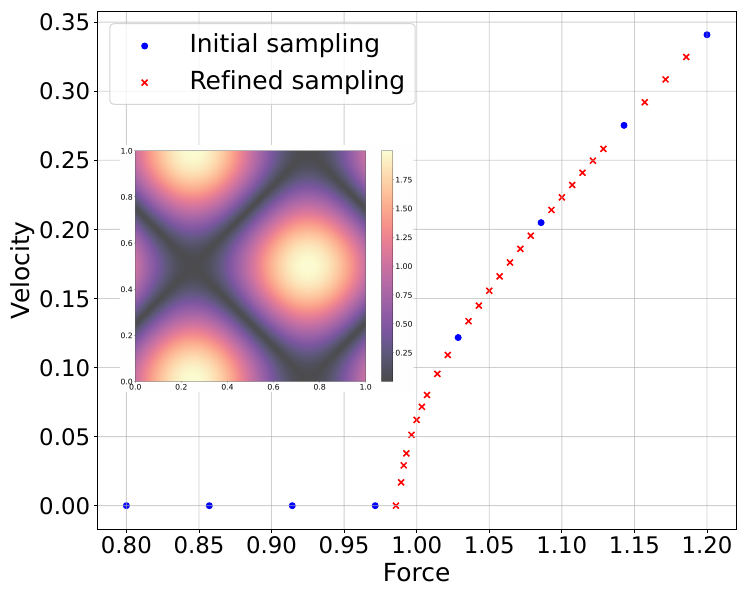}
    \caption{Peeling force vs effective peel front velocity and iterative sampling} 
    \label{fig:iter_sampling_ii}
\end{figure}

We define the {\it effective peel front velocity} to be $v^\text{eff} = \ell/T$.   Further, for a given adhesive pattern $\varphi$ (and peel angle), there is a {\it critical peel force}  $F^*$ below which the front becomes stationary so that $v^\text{eff} = g_\text{per} = 0$ (or equivalently $T=\infty$).  In other words, for a given adhesive pattern, there is a critical force $F^*$ below which it is not possible to peel the film over macroscopic distances, and above which the peel front advances with an effective peel front velocity $v^\text{eff}(F)$ that depends on the peel force.  Further, $v^\text{eff}$ follows a power law near $F^*$, 
\begin{equation} \label{eq:powerlaw}
v^\text{eff} \sim |F-F^*|^\beta \quad F\ge F^*, F \text{ near } F^*
\end{equation} 
with an exponent $0<\beta<1$ that depends on the statistical properties of $\varphi$.  Indeed, for smooth $\varphi$, $\beta = 1/2$.   The relationship $v^\text{eff}$ vs. $F$ describes the overall peeling behavior of the heterogeneous adhesive.  This is shown in Figure \ref{fig:iter_sampling_ii} for the pattern in Figure \ref{fig:peel_front_evo}: note the critical force and the power law behavior near the critical force. 

In this work, we are interested in learning the relationship between the adhesive pattern and the overall peeling behavior, $\varphi \mapsto  F\left(v^\text{eff}\right)$.

%\begin{figure}
%    \centering
%    \includegraphics[width=4in]{fig_dondl_f_v_edit.pdf}
%    \caption{Peeling force vs effective peel front velocity.}
%    \label{fig:dondl_f_v_ii}
%\end{figure}

\subsection{Numerical method}

\paragraph{Fixed force}
Given fixed $F$ and $\varphi$, we  solve the governing equation (\ref{eq:peeling}) for the peel front following Hsueh and Bhattacharya \cite{hsueh_optimizing_2018}.  Ignoring the non-healing or the non-negative constraint for the moment,  it is convenient to take the Fourier transform in $x_1$ and rewrite the peeling equation (\ref{eq:peeling}) in the Fourier domain,
\begin{align} \label{eq:peeling_fourier}
{\hat g}_t (k,t) = F(1-8\pi |k|\hat{g}(k, t)) - \widehat{\varphi(x_{1}, g(x_{1},t))}(k,t), \quad \hat g (k,0) = \hat g_0 (k),
\end{align}
where $ \widehat{\varphi(x_{1}, g(x_{1},t))}$ is the Fourier transform of $f(x_1,t) = \varphi(x_{1}, g(x_{1},t))$.  It is easy to integrate this equation in time by using an explicit forward Euler discretization:
\begin{equation} \label{eq:peeling_fwd_euler}
     \hat{g}^{n+1}(k) = \hat{g}^{n}(k) + \Delta t \left(F(t)(\delta(k) - 8\pi |k|\hat{g}^{n}(k)) -  \widehat{\varphi(x_1, g^{n}(x_1))}(k) \right).
\end{equation}
We have to exercise some care in the computing the nonlinear term.  At the $n^\text{th}$ time-step, we compute $g(x_1,t^n)$ and $f(x_1) = \varphi(x_1,g(x_1,t^n))$, and take the Fourier transform of $f$ to obtain $\hat{f}(k) =  \widehat{\varphi(x_1, g^{n}(x_1))}$.  We then use it to compute the candidate update $ \hat{g}^{n+1}(k)$ using (\ref{eq:peeling_fwd_euler}).

We now reintroduce the no-healing condition.  This has to be imposed in real space,  so we use (\ref{eq:peeling_fwd_euler}) to compute a trial front $\hat{g}^{n+1}_\text{trial}$ in Fourier space and take its inverse Fourier transform to compute a trial front $g^{n+1}_\text{trial}$.  We then set
\begin{equation}
g^{n+1}(x_1) = \max \{ g^{n+1}_\text{trial} (x_1), g^n(x_1) \}.
\end{equation}

\paragraph{Fixed velocity}

Our goal is to study the effective behavior or the relationship $v^\text{eff}(F)$.  However, this relationship is non-smooth at $F^*$, and in fact has an infinite slope at $F^*$ as we saw in the previous section.  So, it is difficult to identify $F^*$ by fixed force calculations (i.e, computing $v^\text{eff}(F)$ and trying to find the largest force where it is zero).  So we follow the approach of Hsueh and Bhattacharya \cite{hsueh_optimizing_2018} to compute $F^*$.  In light of the nature of the effective behavior, it is natural to identify $F^*$ by computing the force as a function of the effective velocity and taking the limit of the effective velocity going to zero.  However, it is hard to find a computational method that imposes an effective velocity.  So, we instead impose the actual (spatially averaged velocity) $\bar{v}$ and find the force $\tilde F(\bar{v},t)$ required to sustain it.  We then define $\bar{F} (\bar{v}) = \max_t \ \tilde F(\bar{v},t)$.  Importantly, $\lim_{\bar{v} \to 0} \bar{F}(\bar{v}) = F^*$.   Further, $\bar{F} (\bar{v})$ behaves similarly to the effective behavior near $(F^*,0)$, and therefore $  F^* \approx \bar{F}(\bar{v})$ for small $\bar{v}$.

We can compute $\tilde F(\bar{v},t)$ and hence $\bar{F}(\bar{v})$ as follows.  Examining (\ref{eq:peeling_fwd_euler}) at $k=0$, we can see 
\begin{align} \label{eq:const_vel_f}
    \tilde F(\bar{v},t) = \bar{v} + \int_0^1\varphi(x_1, g(x_1,t))dx_1,    
\end{align}
so we compute $\tilde{F}(\bar v, t)$ at each iteration and solve for the front evolution by the same forward Euler update in (\ref{eq:peeling_fwd_euler}). 

\begin{figure} 
    \centering
      \includegraphics[width=5in]{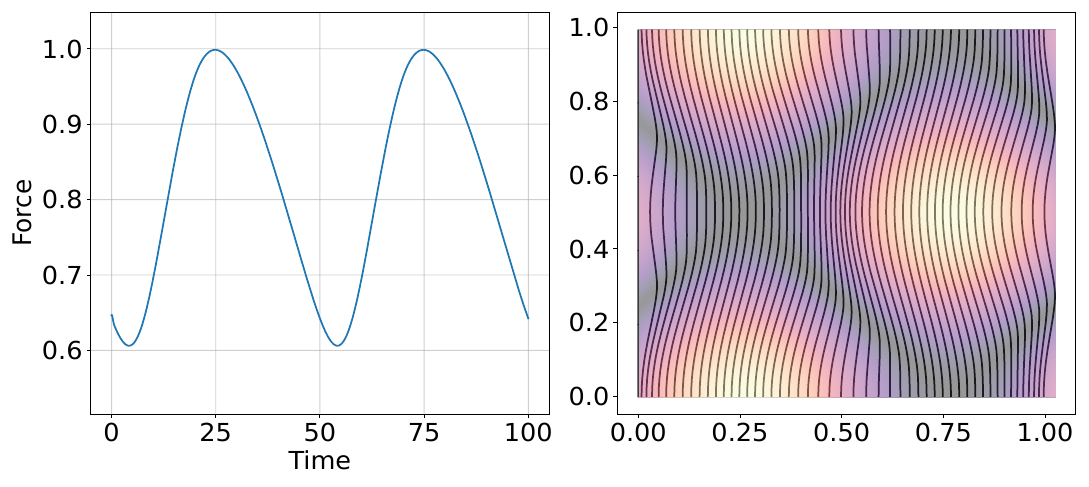}
    \caption{Constant velocity evolution.}
    \label{fig:contour_plot_const_v}
\end{figure}

Figure \ref{fig:contour_plot_const_v} shows the peel front evolution subject to a constant velocity $v_{0} = 0.01$.  In contrast to Figure \ref{fig:peel_front_evo}, here we see a consistent spacing throughout the course of the simulation. The pinning and bending effects are still present, and these manifest themselves through the oscillatory force necessary to sustain this motion.

%%%%%%%%%%%%%%%%%%%%%%%%%%%%%%
%%%%%%%%%%%%%%%%%%%%%%%%%%%%%%
\section{Machine Learning Problems and Methods}
% for Microstructure Characterization
\label{ch:ml_microstructure_characterization}

The goal of this work is to use data generated by repeated numerical simulations to learn the map from the adhesive pattern to the effective adhesive behavior.  
Specifically, we consider two problems.

\begin{itemize}

\item {\bf Problem 1: Critical Force Prediction.}  We seek to learn the map from the adhesive pattern to the critical peeling force, i.e., the map $\varphi \to F^*$.   This is of interest to engineering applications where this critical force determines the loads that one can secure using the adhesive pattern.

\item {\bf Problem 2: Effective Peel Front Velocity Prediction.}  We seek to learn to the map from the adhesive pattern to the critical force, i.e., the map $\varphi \times v^{\text{eff}} \mapsto F$.  %We generate various training samples by running simulations using different constant input forces and computing the corresponding effective velocities.

\end{itemize}

There are a number of challenges in doing so.  We discuss these and the methods we use to overcome them in this section.

%First, $\varphi$ is a function.  It is typically represented by a discretization, but we would like the learnt behavior to be independent of the discretization.  In other words, the map above is an operator and we would like to approximate the operator.  Second, we have to decide the distribution from which we sample our adhesion and forces.

%\subsection{Learning Problems}

%\subsection{Adhesive pattern distribution}

\subsection{Data}

We are interested in smooth adhesive patterns and therefore sample our patterns to be the absolute value of a sum of sinusoids with amplitudes drawn from a normal distribution:
\begin{equation} \label{eq:phi_para}
{\mathcal D} = \left\{    \varphi(x_{1}, x_{2}) = \left| \frac{1}{\sum \left| a_n \right|} \sum_n^N \left| a_n \right| \sin(2 \pi n x_{1})\cos(2 \pi n x_{2})  \right|, a_n \in \mathcal{N}(0, 1) \right\} .
\end{equation}
We take $N=128$.  Note that each pattern is normalized so that $\varphi \le 1$. We generate data for Problem 1 by  sampling various microstructures from the distribution (\ref{eq:phi_para}) and computing the critical peeling forces using fixed velocity simulations. 

%\begin{figure}
%    \centering
%    \includegraphics[width=4in]{g11049.pdf}
%    \caption{Iterative sampling} 
%    \label{fig:iter_sampling_ii}
%\end{figure}

For Problem 2, we sample microstructures from the distribution (\ref{eq:phi_para}), and then sample the peel force for each microstructure.  This latter problem requires care in light of the nature of the peel force vs.\ effective peel front relationship that is non-smooth with a power law at the critical force (cf. (\ref{eq:powerlaw}) and Figure \ref{fig:iter_sampling_ii}).  If input force values were uniformly sampled from a given range, we risk undersampling the depinning regime. To mitigate this, we apply an iterative sampling algorithm: we first sample at fixed intervals of the peel force, and then iterate by bisection (i.e., adding sampling points at the midpoint between two adjacent points) until the  adjacent velocities $v^\text{eff}$ values are bounded from above by a prescribed value.   Figure \ref{fig:iter_sampling_ii} shows the initial and final sampling: note that there is a higher concentration of samples around the depinning regime due to the steeper gradient.

\subsection{Neural Approximation} \label{ssec:nn_model_selection}

\begin{figure}[t]
    \centering
    \includegraphics[width=6in]{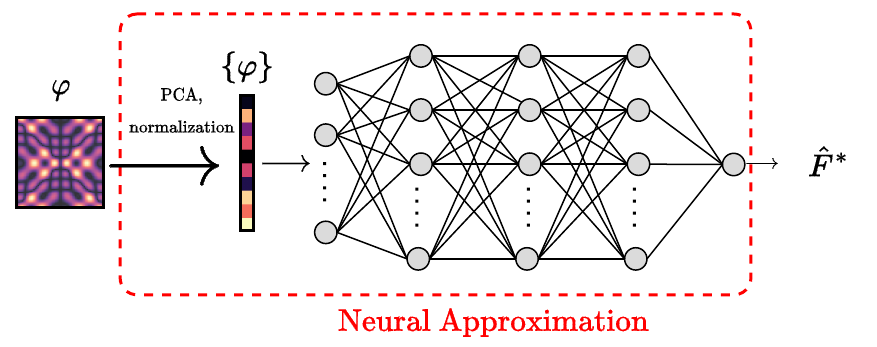}
    \caption{Neural approximation combining model reduction and deep neural network}
    \label{fig:training_schematic}
\end{figure}

An important challenge in these problems is that the adhesive pattern $\varphi$ is a function.  It is typically represented by a discretization, but we would like the learnt behavior to be independent of the discretization.  In other words, the map we seek to approximate is an operator.   We use an approach following the PCA-Net architecture of Bhattacharya {\it et al.} \cite{bhattacharya_2021} that combines model reduction and neural networks.  The key idea is to approximate the map using a composition of a projection of the function to a lower dimensional set using principal component analysis (PCA) and a deep neural network:
$$ y = (NN \circ P) X $$
where $X$ is the input, $y$ the output, $P$  a finite-dimensional projection and $NN$ a deep neural network.   This is shown schematically in Figure \ref{fig:training_schematic} for Problem 1.

Consider Problem 1.  Our input (adhesive pattern) $X \in \mathbb{R}^{M^2}$ where $M$ is the grid resolution at which the data is provided.  PCA identifies a $K$ dimensional sub-space with $K<<M^2$ and projects the input to that subspace.  Specifically, (after some shifting if necessary), PCA provides a basis set $\{X^k\}_{k=1}^K$ so that PCA is the operation $x=PX$ where 
$$
P = \sum_{k=1}^K X^k \otimes X^k 
$$
Now, $x$ is used as an input for a deep neural network.  Thus, our neural approximation may be written as
\begin{gather}
y = B x^{L}, \nonumber \\
x^{(l)} = \sigma(W^{(l)}x^{(l - 1)} + b^{(l)}), \quad l = 1 \hdots L, \label{eq:arch} \\
x^{(0)} = P X, \nonumber
\end{gather}
where $\{W^{(l)}\}_{l=1}^L$ are weights and $\{b^{(l)}\}_{l=1}^L$ are biases to be identified by minimizing a loss function to be chosen with respect to the data \cite{MURTAGH1991183}, and $\sigma$ is an activation function (taken to be ReLU \cite{10.5555/3104322.3104425}).

In Problem 2, our input is the image and the effective peel front velocity $X = \varphi \oplus v^\text{eff} \in \mathbb{R}^{M^2 + 1}$.  PCA is used only in the image so that we may replace the last line of (\ref{eq:arch}) with
\begin{equation}
x^{(0)} = P X \oplus v^\text{eff}.
\end{equation}

In both problems, we first train the PCA, and then use its output to train the neural network.   The projected PCA dimension is $54$ so that the explained variance is at least $0.99$.  We apply a grid search approach for training the hyperparameters in the neural network. This means we test every possible combination of the hyperparameters of interest in a brute-force manner. The Adam optimizer \cite{https://doi.org/10.48550/arxiv.1412.6980} is used, in addition to early stopping which stops the training if the validation loss has not improved in $50$ epochs.  The hyperparameter space is indicated in Table \ref{tab:exp-1-2-hyperparams}.   More sophisticated alternatives include randomized grid search or Bayesian optimization, but in the context of this work, the brute-force approach is tractable. 

\begin{table}
\centering
\caption{Hyperparameter space for critical force learning problem}
\label{tab:exp-1-2-hyperparams}
\begin{tabular}{ll}
\hline
\rowcolor[HTML]{EFEFEF} 
Hyperparameters        & Tested Values                 \\ \hline
Hidden Layers          & {[}2, 4, 8{]}                 \\
Hidden Layer Dimension & {[}64, 128, 256, 512, 1024{]} \\
Batch Normalization    & {[}True, False{]}             \\
Dropout                & {[}0.0, 0.2, 0.5{]}           \\
Loss Functions         & {[}L1\_loss, MSE\_loss{]}     \\
Learning Rate          & {[}0.001, 0.01, 0.1{]}        \\
Number of epochs       & {[}1000{]}                    \\
Batch Size             & {[}128, 256{]}                \\ \hline
\end{tabular}
\end{table}

%%%%%%%%%%%%%%%%%%%%%%%%%%%%%%
%%%%%%%%%%%%%%%%%%%%%%%%%%%%%%
\section{Learning Effective Adhesion}
\label{ch:results_characterization}

In this section we present the results of the  approach proposed in Section \ref{ch:ml_microstructure_characterization} concerning the characterization of the effective adhesive behavior using neural networks.

\subsection{Problem 1: Critical Force Predictions} \label{sec:nn_problem_1_results}

\begin{figure}
    \centering
    \begin{subfigure}[t]{0.4\textwidth}
    \includegraphics[width=2.5in]{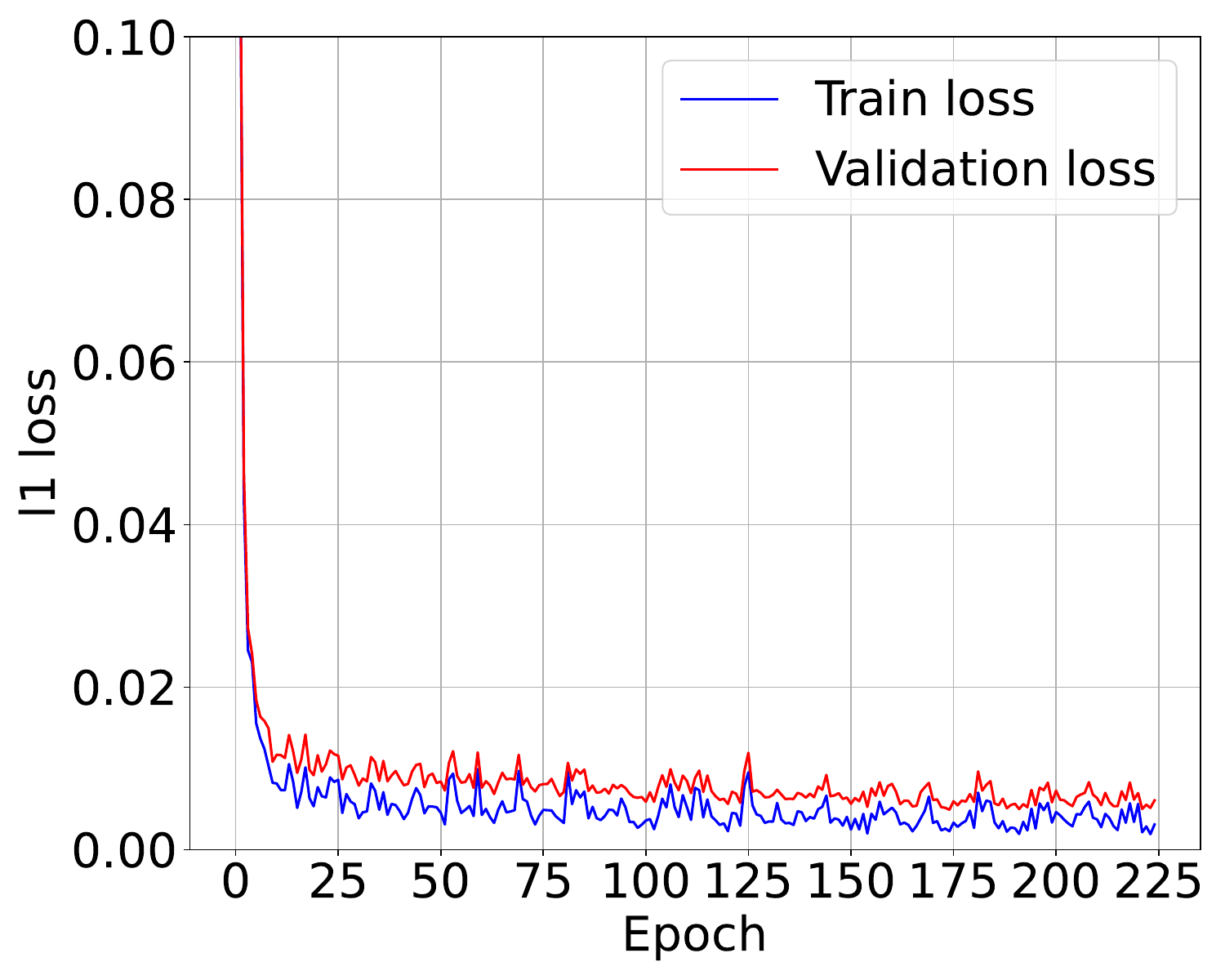} 
    \caption{L1 loss for Problem 1}
    \end{subfigure}
    \begin{subfigure}[t]{0.4\textwidth}
          \includegraphics[width=2.5in]{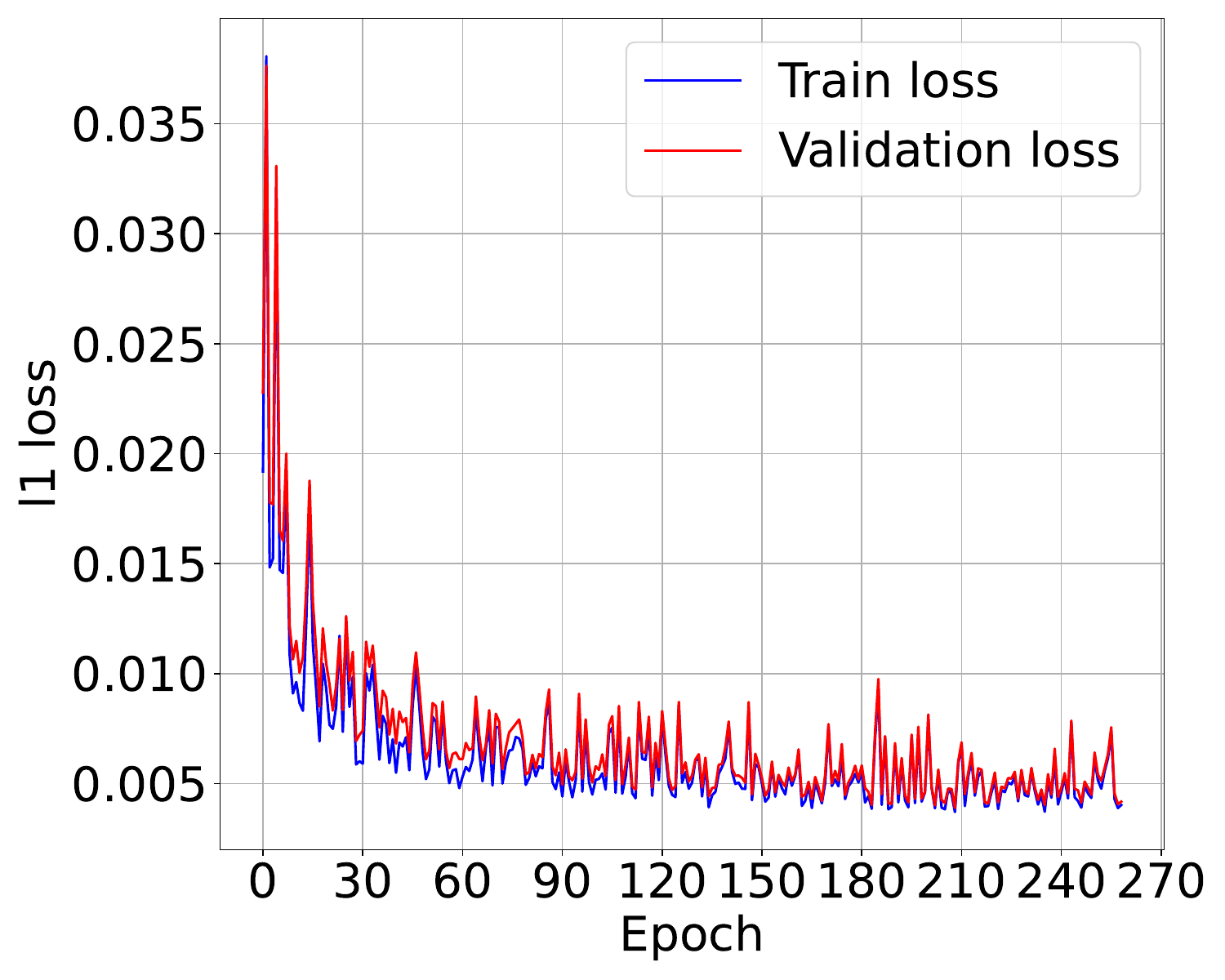}
    \caption{L1 loss for Problem 2}
    \end{subfigure} 
    \caption{Training and validation loss per epoch for (a) Problem 1 and (b) Problem 2.} 
    \label{fig:loss}
\end{figure}

The data for Problem 1 comprises $1024$ training samples, $256$ validation samples, and $256$ test samples. The microstructures are different across all splits (none are reused in any split), so there are $1,536$ unique microstructures in total.  Table \ref{tab:param} shows the training, validation, and test losses of the best performing models according to the validation loss metric, as well as their corresponding hyperparameters (for both of the tested loss functions). Figure \ref{fig:loss}(a) shows the loss curves for the training and validation data for the model trained using the $L1$ loss.

\begin{table}
\centering
\caption{Tuned model configurations and losses.}
%\hl{Why these cases?  E.g., why 2 layers for L1 and 4 for MSE in Problem 1}}
\label{tab:param}
\resizebox{\textwidth}{!}{\begin{tabular}{l|l|l|l|l|l|l|l|l|l|l}
\textbf{Case} & \textbf{Nodes} & \textbf{L} & \textbf{BN} & \textbf{D} & \textbf{Loss} & \textbf{LR} & \textbf{Batch} & \textbf{Train Loss} & \textbf{Val Loss} & \textbf{Test Loss} \\ \hline
\multicolumn{11}{c}{Problem 1: Critical Force}\\
P1-L1 &   $1024$ & $2$ & False & $0.0$ & L1 & $10^{-3}$ & $128$ & $2.62 \cdot 10^{-3}$ & $4.93 \cdot 10^{-3}$ & $4.53 \cdot 10^{-3}$ \\
P1-MSE   & $1024$ & $4$ & False & $0.0$ & MSE & $10^{-3}$ & $128$ & $2.35 \cdot 10^{-7}$ & $3.76 \cdot 10^{-5}$ & $2.64 \cdot 10^{-5}$\\
\hline
\multicolumn{11}{c}{Problem 2: Effective Peel Front Velocity}\\
%P2-L1 &   $128$ & $2$ & False & $0.0$ & L1 & $10^{-2}$ & $128$ & $6.15 \cdot 10^{-3}$ & $1.09 \cdot 10^{-2}$ & $9.94 \cdot 10^{-3}$ \\
%P2-MSE   &    $1024$ & $8$ & False & $0.0$ & MSE & $10^{-3}$ & $256$ & $8.51 \cdot 10^{-5}$ & $2.72 \cdot 10^{-4}$ & $3.48 \cdot 10^{-4}$\\
P2-L1 &   $1024$ & $2$ & False & $0.0$ & L1 & $10^{-2}$ & $256$ & $4.02 \cdot 10^{-3}$ & $4.18 \cdot 10^{-3}$ & $4.18 \cdot 10^{-3}$ \\
P2-MSE   &    $128$ & $2$ & False & $0.0$ & MSE & $10^{-2}$ & $128$ & $9.27 \cdot 10^{-5}$ & $1.02 \cdot 10^{-4}$ & $1.01 \cdot 10^{-4}$\\
\hline
\multicolumn{11}{c}{L: Layers, BN: Batch Normalization, D: Dropout}\\
\end{tabular}}
\end{table}

\begin{figure}
    \centering
    \includegraphics[width=3in]{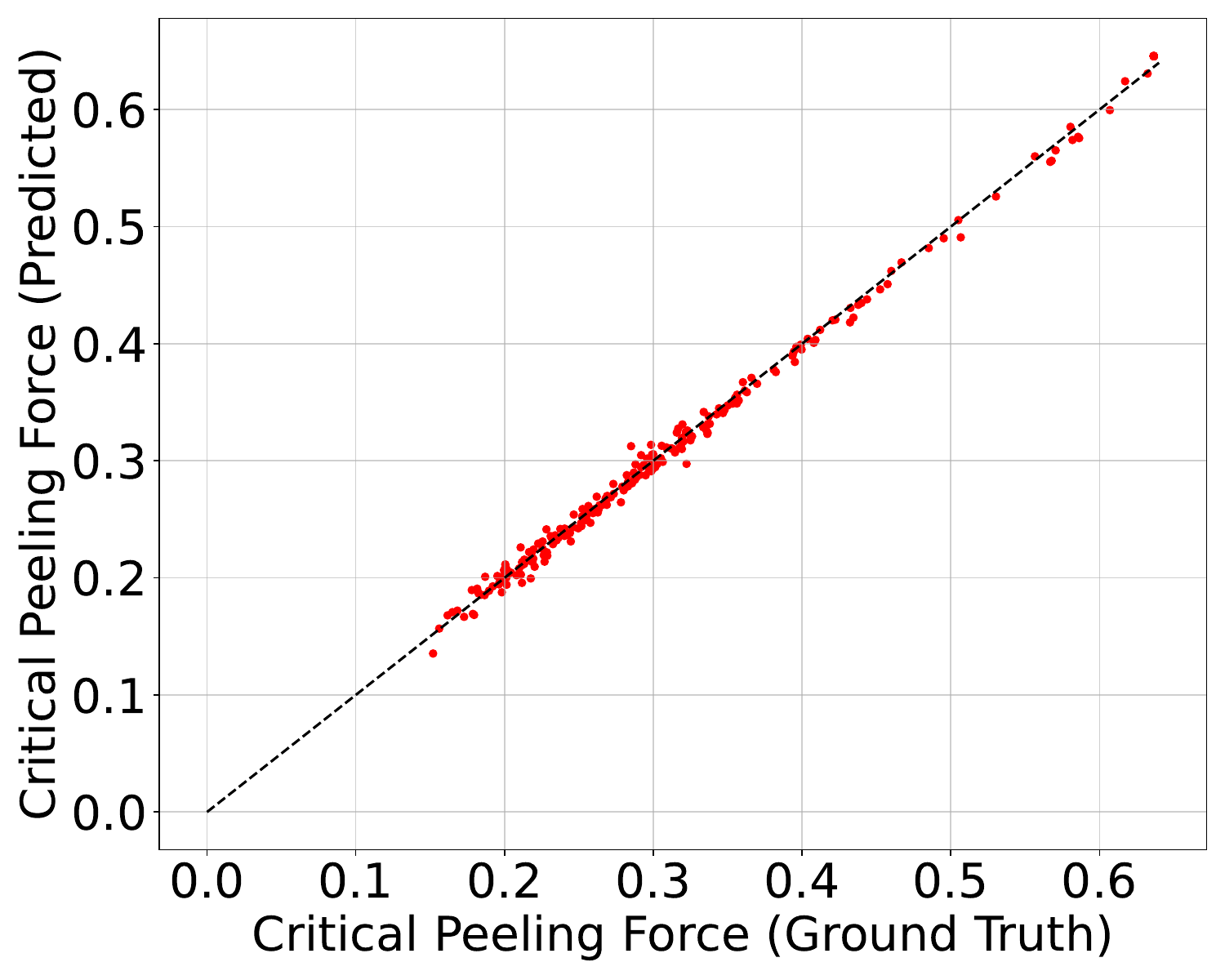}
    \caption{Predicted vs ground truth critical forces}
    \label{fig:f_crit_scatter}
\end{figure}

To get a more concrete idea of the performance, we consider the relative absolute test error of the $k$-th sample, denoted by 
\begin{equation} \label{eq:rel_abs_err}
{\mathcal E}_{k} = \frac{|F^{*}_{k} - \hat{F}^{*}_{k}|}{F^{*}_{k}},
\end{equation}
where $\hat{F}^*_k$ is the model prediction. Table \ref{tab:rel_errors} shows the performance in terms of this metric (maximum, minimum and average over $k$) for the test data.   
In both the case of the L1 and MSE loss functions we see that the models have learned to predict the effective adhesive strength for previously unseen microstructures with an average error of roughly $1.8\%$.  Figure \ref{fig:f_crit_scatter} shows a scatter plot of the predicted and ground-truth critical forces for all the test samples predicted by the model trained using the L1 loss function, showing good general agreement across the entire regime.

\begin{table}
\centering
\caption{Relative absolute errors}
\label{tab:rel_errors}
\begin{tabular}{l|l|l|l}
\textbf{Case} &  \textbf{Min.\ Error}  &  \textbf{Max.\ Error}  &  \textbf{Ave.\ Error}  \\ \hline
\multicolumn{4}{c}{Problem 1: Critical Force}\\
P1-L1 & $9.98 \cdot 10^{-5}$ & $1.09 \cdot 10^{-1}$ & $1.86 \cdot 10^{-2}$ \\
P1-MSE & $1.00 \cdot 10^{-5}$ & $1.45 \cdot 10^{-1}$ & $1.79 \cdot 10^{-2}$ \\
\hline
\multicolumn{4}{c}{Problem 2: Effective Peel Front Velocity}\\
% P2-L1 & $2.01 \cdot 10^{-5}$ & $2.70 \cdot 10^{-1}$ & $2.46 \cdot 10^{-2}$ \\
% P2-MSE & $8.53 \cdot 10^{-6}$ & $3.85 \cdot 10^{-1}$ & $2.94 \cdot 10^{-2}$              
P2-L1 & $4.71 \cdot 10^{-6}$ & $2.24 \cdot 10^{-1}$ & $1.16 \cdot 10^{-2}$ \\
P2-MSE & $3.63 \cdot 10^{-7}$ & $2.19 \cdot 10^{-1}$ & $1.31 \cdot 10^{-2}$                  
\end{tabular}
\end{table}

\subsection{Problem 2: Force vs. Velocity Predictions}
%$3989$, $997$, and $991$
% $128$, $32$, and $32$
The data for problem 2 consists of $31,986$, $7,949$, $7,991$ train, validation, and test samples respectively, drawn from  $1024$, $256$, $256$ distinct microstructures.  Recall that we sample multiple forces for each microstructure. The resulting $v^\text{eff}$ values from the simulations lie between $[0.0, 0.67]$.  Table \ref{tab:param} shows the best performing models according to their validation losses and Figure \ref{fig:loss}(b) shows the L1 loss for both the training and test data.

\begin{figure}
    \centering
    \begin{subfigure}[t]{0.4\textwidth}
        \centering
        \includegraphics[width=3in]{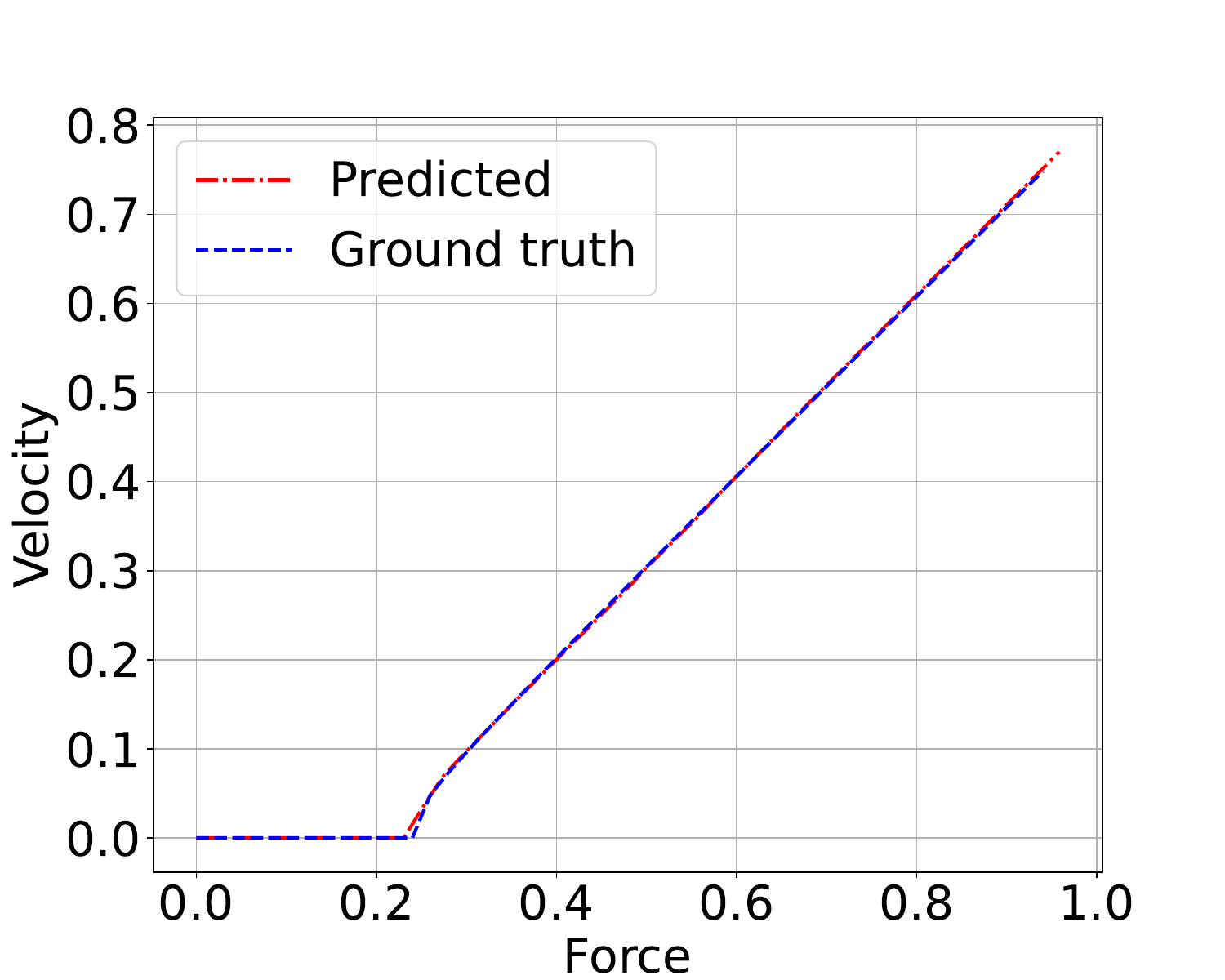}
        \caption{Sample $40$} 
        \label{fig:f-v_40}
    \end{subfigure}%
    \hfill
    \begin{subfigure}[t]{0.4\textwidth}
        \centering
        \includegraphics[width=3in]{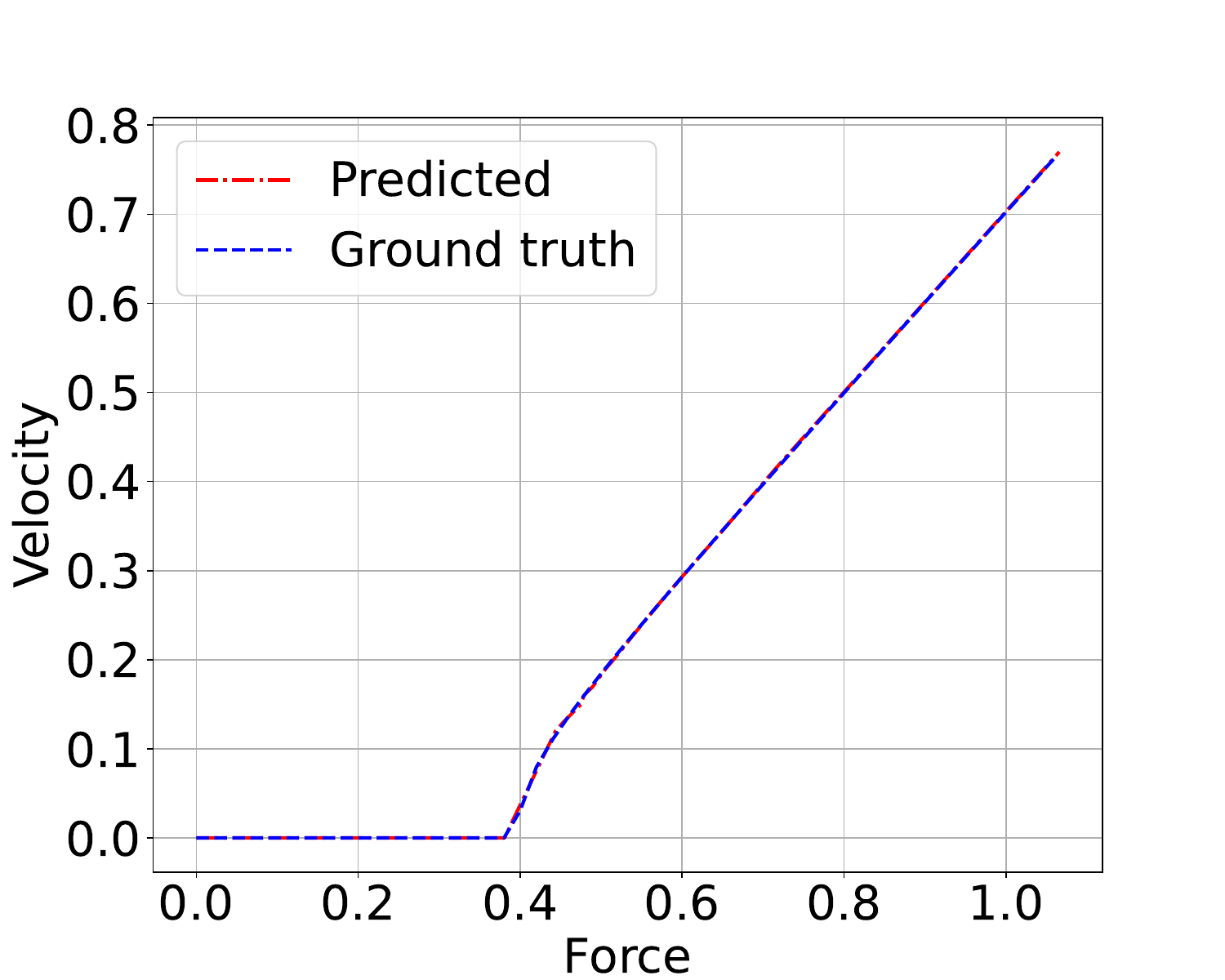}
        \caption{Sample $70$}
        \label{fig:f-v_70}
    \end{subfigure}
    \caption{Ground truth vs. neural approximation for force-velocity}
    \label{fig:f-v_plots}
\end{figure}

The relative absolute errors for all test samples (including those that were pinned), as given by (\ref{eq:rel_abs_err}), is shown in Table \ref{tab:rel_errors}.  We see that we we achieve a test error of $1.16\%$ and $1.31\%$ for the L1 and MSE models respectively, which can be further illustrated by Figure \ref{fig:f-v_plots} which shows the predictive capacity of the L1 model across a broad range of the force-velocity regime. 

%%%%%%%%%%%%%%%%%%%%%%%%%%%%%%
%%%%%%%%%%%%%%%%%%%%%%%%%%%%%%
\section{Conclusion} \label{sec:conclusion}

In this paper, we have proposed and demonstrated a neural approximation that is able to learn the effective adhesive properties experienced by a thin film being peeled from a heterogenous surface.  In doing so, we addressed two challenges.  The first is the fact that we want the neural approximation to be independent of the discretization that is used to describe the adhesive pattern.  The second is harder.  The evolution of the peel front is characterized by pinning and de-pinning.  Consequently, there is a critical force below which there is no peeling, and above which the peel front advances with an effective peel velocity.  Further, the relationship between the force vs. peel front velocity is singular near the critical force.  

While this work addressed adhesion, the ideas and the approach are applicable to a variety of other free boundary problems.  It also provides a path forward for a free discontinuity problem like fracture.  If we know the surface on which the crack will propagate, as for example in important problems like composite materials or interfacial fracture, then the approach proposed here is readily extendible.  In fact, the equations governing the propagation of an almost planar crack through a heterogeneous medium is very similar to (\ref{eq:peeling}) \cite{gao_first-order_1989}.   Thus, our approach will carry over.  The situation where the crack deviates significantly from a planar crack or branches is difficult and the topic of future research.

%%%%%%%%%%%%%%%%%%%%%%%%%%%%%%
%%%%%%%%%%%%%%%%%%%%%%%%%%%%%%
\section{Acknowledgements}

We are delighted to acknowledge  helpful discussions with Professor Jean-Fran\c{c}ois Molinari.  We gratefully acknowledge the financial support of the Army Research Office through grant number W911NF-22-1-0269.  The simulations reported here were conducted on the Resnick High Performance Computing Cluster at the California Institute of Technology.

%Bibliography

\end{document}